\documentclass[showpacs,preprintnumbers]{revtex4}
\usepackage{amssymb}
\usepackage{amsmath}
\usepackage{graphicx}
\usepackage{dcolumn}
\usepackage{bm}
\usepackage{epsf}

\begin{document}

\preprint{APS/123-QED}
\title{Testing Non-local Nucleon-Nucleon Interactions in the Four-Nucleon Systems}
\author{Rimantas Lazauskas}
\affiliation{CEA/DAM/DPTA Service de Physique Nucl\'eaire, BP 12, F-91680 Bruy\`eres-le-Ch\^atel, France.}
\author{Jaume Carbonell}
\affiliation{Laboratoire de Physique Subatomique et de Cosmologie, 53 Avenue des
Martyrs, 38026 Grenoble Cedex, France.}
\date{\today }

\begin{abstract}
The Faddeev-Yakubovski equations are solved in configuration space for the $%
\alpha$-particle and n-$^3$H continuum states. We test the ability of
nonlocal nucleon-nucleon interaction models to describe 3N and 4N systems.
\end{abstract}

\maketitle

\section{Introduction}

\label{sec:level1}

In a fundamental description of nucleon-nucleon (NN) interaction,
the existence of the nucleon internal structure can not be ignored.
The standard NN
potentials are actually effective tools aiming to mimic a much more
complicated interaction process, of which it is not even clear that it
could be reduced to a potential problem. The NN system can be rigorously described only when
starting from the underlying QCD theory for the nucleon constituents: quarks
and gluons. This is however a very difficult task, which is just becoming
accessible in lattice calculations \cite{BBPS_PLB585,Beane_FB17_03,WPWG}, and that will be in any
case limited for a long time to the two-nucleon system. In any attempt to
describe the nuclear structure, one is thus obliged to rely on more or less
phenomenological models.

Since the nucleon size is comparable to the strong interaction range, the
effects of its internal structure are expected to be considerable.
In particular the NN interaction should be non-local, at least for small
inter-nucleon distances.
In addition, we have no reason to believe that
nuclear interaction is additive as the Coulomb one:
the interaction between
two nucleons may not be independent of the presence of a third one
in their vicinity. Finally the interplay of nucleon confinement and relatively
large kinetic energies can generate -- e.g. via virtual nucleon excitations -- a
rather strong energy-dependence in the interaction.
Despite numerous studies devoted to this subject, we still do not have a
clear understanding of the relative importance of these effects in NN force,
specially concerning their influence on experimentally measurable quantities.

\begin{figure}[h!]
\begin{center}
\mbox{\epsfxsize=10.0cm\epsffile{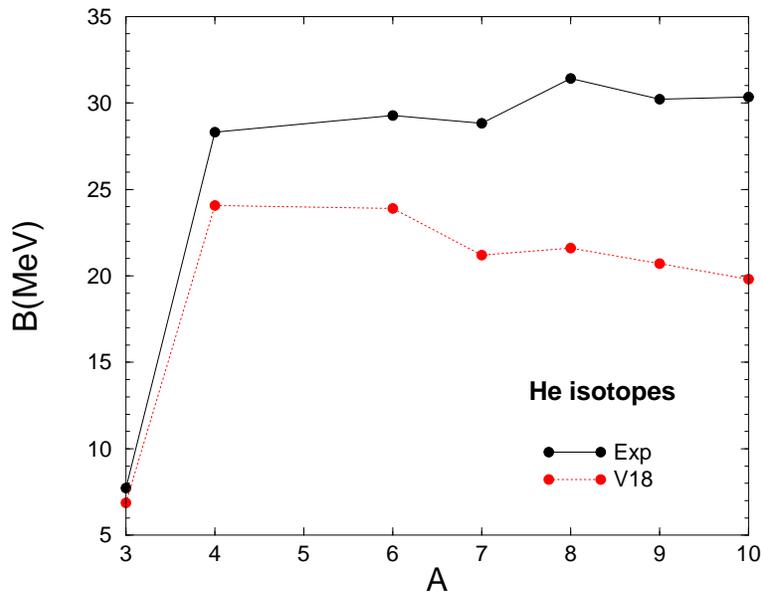}}\hspace{1.cm}
\end{center}
\caption{"(Color online)" Comparison between the experimental and theoretical calculations
with Argonne v18 interaction in He isotopes. Results are taken from \protect\cite{PPWC_PRC64_01,PVW_PRC66_02}.} \label{Fig_He_Exp_V18}
\end{figure}

This work investigates the consequences of using non-local NN forces in describing
the A=3 and A=4 nuclear systems.
The locality of NN force,  assumed in some of the so called realistic models \cite{AV18_PRC_95,NIJ_PRC49_94},
is  due more to numerical convenience than to convincing physical arguments.
The two-nucleon experimental data, since they contain only on-shell physics,
are successfully reproduced without including any energy dependence or
non-locality in the NN force. However, they all suffer from the
underbinding problem, i.e.  two-nucleon interaction alone fails to
reproduce the nuclear binding energies, starting already from the simplest
A=3 nuclei. Figure \ref{Fig_He_Exp_V18} shows the relative differences
between experimental and theoretical binding energies for He isotopes
obtained with AV18 potential \cite{AV18_PRC_95,PPWC_PRC64_01,PVW_PRC66_02}.
These differences increase with the mass number A and vary from  $\sim 0.7$ MeV
in $^3$He to $\sim 10$  MeV in the case of $^{10}$He.

The inclusion of non-local terms -- like in Nijm 93 and Nijm I potentials
\cite{NIJ_PRC49_94} --
does not remove this discrepancy \cite{FPSdS_93,Nogga_alp}.
If in some cases, like in CD-Bonn \cite{CD_Bonn} or in chiral models
\cite{EM_PLB524_02,EGM_NPA637_98,EGM_NPA671_00},
they considerably improve 3- and 4-N binding
energies, the improvement is still not sufficient to reproduce the experimental values.
This underbinding is rather easily removed by means of three-nucleon forces (3NF).
The existence of such forces is doubtless, but their strength
depends on the NN partner in use and is  determined only by  fitting requirements.

However the use of 3NF, to some extent, can be just a matter of taste. It
has been shown in \cite{GP_FBS9_90, Polyzou_PRC58_98}
that two different, but phase-equivalent, two-body interactions
are related by a unitary, non-local, transformation.
One thus could expect that a substantial part of 3- and multi-nucleon forces could
also be absorbed by non local terms.
A considerable simplification would result if
the bulk of experimental data could be described by
only using two-body non-local interaction.
In fact, the unique aim of any phenomenological  model is
to provide a satisfactory description of the experimental observables
but it is worth reaching this aim by using the simplest possible approach.

There exist already calculations reproducing the triton binding energy without
making explicit use of 3NF. The first one was obtained by Gross and Stadler \cite{Gross}
using a relativistic equation and  benefiting from some additional freedom
in the off-shell vertex form factors. The complexity in using relativistic
equations makes however difficult their extension to larger nuclear systems, or
even to 3N scattering.

A very promising result, which takes profit from non-locality in
non relativistic nuclear models, has been obtained by Doleschall and collaborators
\cite{D_NPA602_96,D_FBS_98,DB_FBS_99,DB_PRC62_00,DB_NPA684_01,DBPP_PRC67_03,D_PRC_04}.
In this series of papers, purely phenomenological non-local NN forces
have been constructed, which were able to overcome the lack of binding energy
in three-nucleon systems, namely $^{3}$H and $^{3}$He, without explicitly
using 3NF and still reproducing 2N observables. The striking success of these
models is closely related to the presence of small deuteron D-state probability,
comparatively to local NN interaction.
A difference which does not contradict the phenomenology
and which follows from using two equivalent representations
of the one and the same physical object \cite{AD_NPA714_03}.
In fact, non-locality of the
Doleschall potential softens the short range repulsion of local NN
models and can simulate part of effects due to the 6-quark structures as well as
quark exchange between  nucleons inside the nucleus.
Local realistic interaction models, maybe with exception of Moscow \cite{Moscou}
potential, artificially prohibits such effects by imposing a very strong  short-range repulsion.

Our work is an extension to the A=4
nuclei of the Doleschall pioneering studies.
In particular, we would like to check whether such non-local
interaction models remain successful or not when applied to the more complicated
4N systems. We will first provide results for $^4$He ground state
and then extend the calculations to the n-$^{3}$H scattering. This
system possesses a resonance at E$_{cm}\approx 3$ MeV, exhibiting different
dynamical properties from those of bound states and testifying a failure of
the conventional NN+3NF interaction models \cite{These_Rimas_04}.

\section{Theoretical treatment}

\subsection{FY\ Equations \label{sec_FY_eq}}

\begin{figure}[h!]
\begin{center}
\mbox{\epsfxsize=7.cm\epsffile{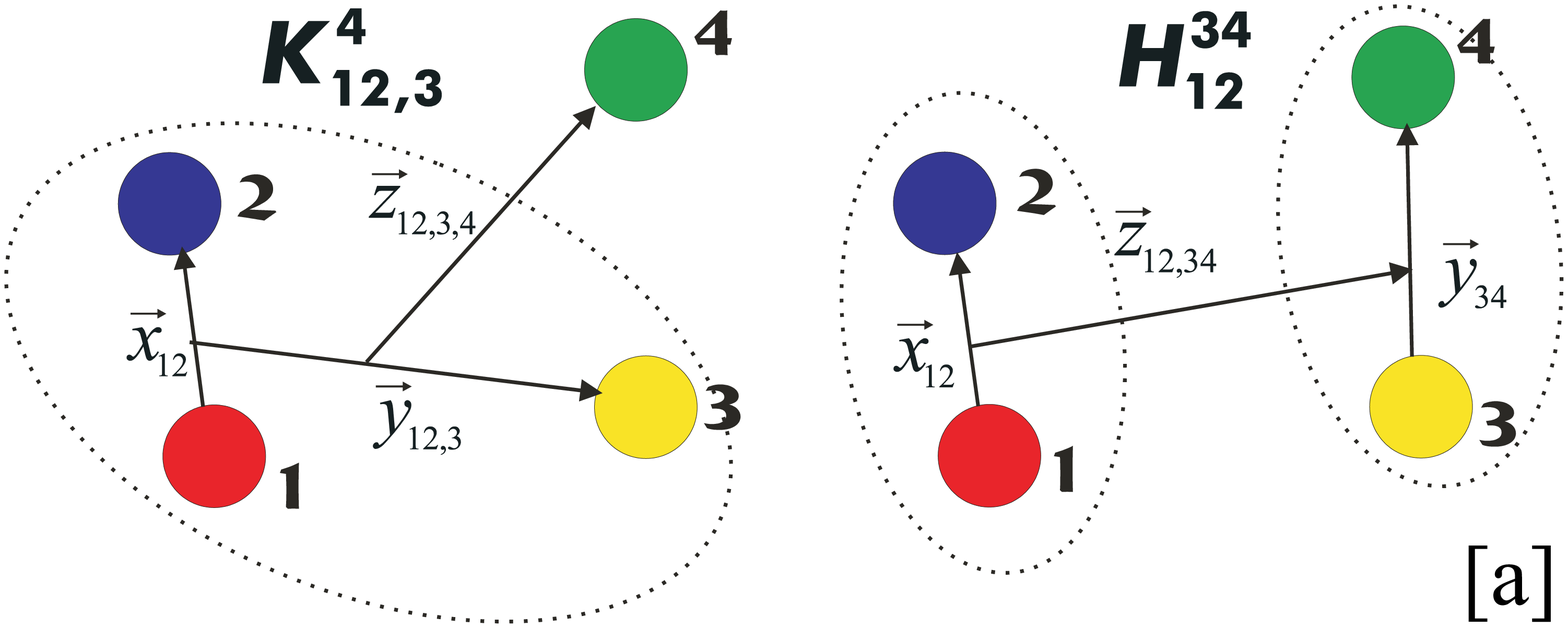}} \hspace{2.cm} %
\mbox{\epsfxsize=7.cm\epsffile{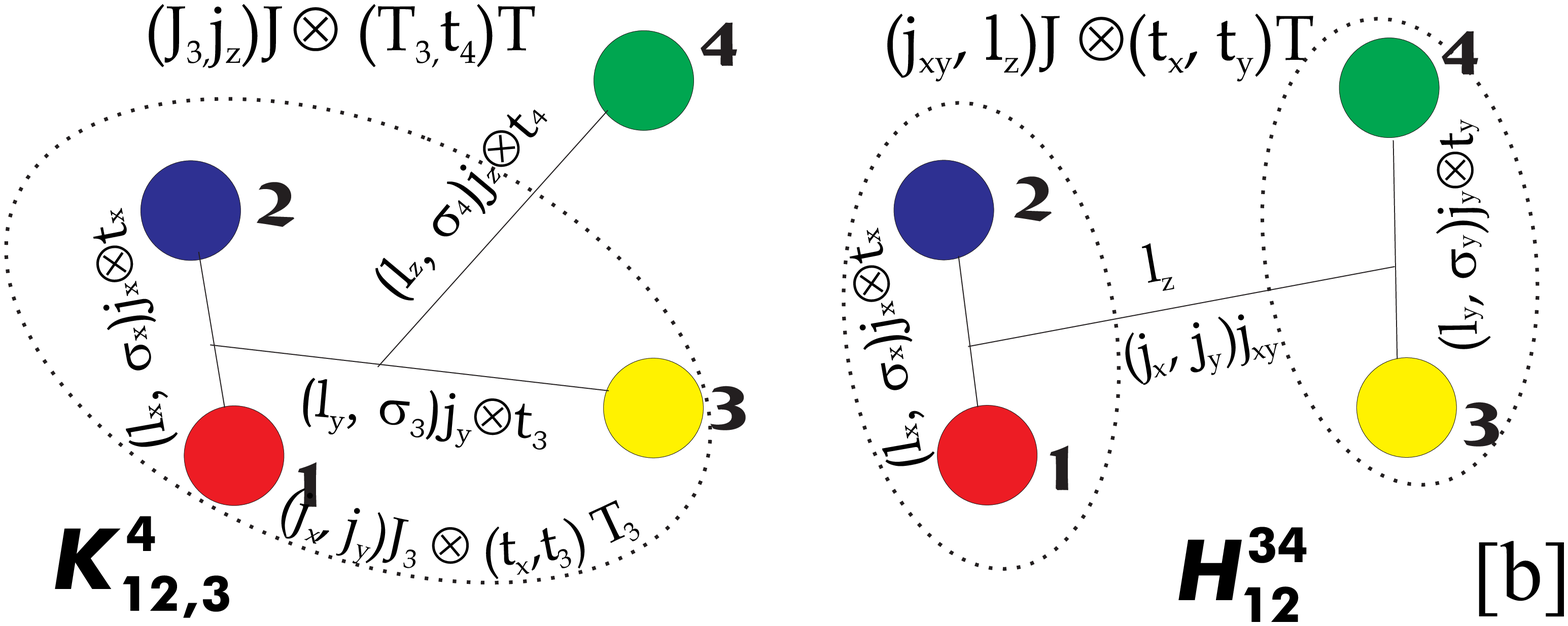}}
\end{center}
\caption{"(Color online)" Figure [a]: FY components $K$ and $H$.
Asymptotically, as $z\rightarrow \infty$,
components K describe 3+1 particle channels, whereas components H contain
asymptotic states of 2+2 channels. Figure [b] illustrates the
j-j coupling schemes used when developing FYC $K$ and $H$ into partial wave basis.}
\label{Fig_4b_config}
\end{figure}

We describe the 4N\ system by using Faddeev-Yakubovski (FY)
equations in configuration space \cite{Fadd_art,Yakub,Merk_TMP56}.
Even though the major goal of FY formalism is a mathematically
rigorous description of the continuum states, it turns out as well
to be advantageous when dealing with bound state problem. The
advantage lies in the natural decomposition  of the wave function
in terms of the so called Faddeev-Yakubovski components (FYC)
which take benefit of the systems symmetry properties. These
amplitudes have simpler structure than the wave function itself
and are therefore easier to handle numerically. Four-particle
systems require the use of two types of FYC, namely $K$ and $H$.
Asymptotes of components $K$  incorporate 3+1 (see
Fig.\ref{Fig_4b_config}) channels, while components H contain
asymptotes of 2+2 ones
 (see Fig.\ref{Fig_4b_config}). By permuting the particles
one can construct twelve different components of the type $K$ and six
components of the type $H$. The total wave function is simply a sum of these 18 FYC.

In theoretical treatment of the problem we use the isospin formalism, i.e.
we consider protons and neutrons as being degenerate states of the same
particle - nucleon. For a system of identical particles, the  FYC are
not completely independent, being related by  several straightforward
symmetry relations. All the 18 FYC can be obtained by the action of the
permutation operators on two of them, arbitrarily chosen, provided one is of
type $K$ and the other of type $H$. We have chosen $K\equiv \Psi _{12,3}^{4}$
and $H\equiv \Psi _{12}^{34}.$ The four-body problem is solved by
determining these two components, which satisfy the system of differential equations%
\cite{Merk_TMP56}:
\begin{eqnarray}
\left( E-H_{0}-V\right) K &=&V(P^{+}+P^{-})\left[ (1+Q)K+H\right]  \notag \\
\left( E-H_{0}-V\right) H &=&V_{12}\tilde{P}\left[ (1+Q)K+H\right] .
\label{FY1}
\end{eqnarray}%
$P^{+}$, $P^{-}$, $\tilde{P}$ and $Q$ are the permutation operators:
\begin{eqnarray*}
P^{+}&=&(P^{-})^{-1}=P_{23}P_{12}\cr Q&=&\varepsilon P_{34}\cr \tilde{P}%
&=&P_{13}P_{24}=P_{24}P_{13}.
\end{eqnarray*}
Employing operators defined above, systems wave function is given by
\begin{equation}  \label{FY_wave_func}
\Psi =\left[ 1+(1+P^{+}+P^{-})Q\right] (1+P^{+}+P^{-})K +(1+P^{+}+P^{-})(1+%
\tilde{P})H.
\end{equation}

Components $K$ and $H$ are functions in configuration space, and
depend also on the internal degrees of freedom of the individual
particles (spins and isospins). The configuration space is provided by the
position of the different particles, which we describe by using reduced
relative coordinates. These coordinates differ from Jacobi coordinates
usually employed in Classical Mechanics by factors depending on the particle
masses. Use of such coordinates has several big advantages: first center of
mass motion can be easily separated, then transition between two bases is
equivalent to orthogonal transformation in $R^{3(N-1)}$ space; finally,
kinetic energy operator in this basis reduces to multidimensional Laplace
operator in corresponding subspaces. Two principally different sets of
reduced relative coordinates can be defined. One is associated with the
components $K\equiv \Psi _{ij,k}^{l}$:

\begin{equation}
\begin{tabular}{ll}
$\overrightarrow{x_{ij}}$ & $=\sqrt{2\frac{m_{i}m_{j}}{m_{i}+m_{j}}}(%
\overrightarrow{r}_{j}-\overrightarrow{r}_{i})\smallskip $ \\
$\overrightarrow{y_{ij,k}}$ & $=\sqrt{2\frac{\left( m_{i}+m_{j}\right) m_{k}%
}{m_{i}+m_{j}+m_{k}}}(\overrightarrow{r}_{k}-\frac{m_{i}\overrightarrow{r}%
_{i}+m_{j}\overrightarrow{r}_{j}}{m_{i}+m_{j}})\smallskip $ \\
$\overrightarrow{z_{ijk,l}}$ & $=\sqrt{2\frac{\left(
m_{i}+m_{j}+m_{k}\right) m_{l}}{m_{i}+m_{j}+m_{k}+m_{l}}}(\overrightarrow{r}%
_{l}-\frac{m_{i}\overrightarrow{r}_{i}+m_{j}\overrightarrow{r}_{j}+m_{k}%
\overrightarrow{r}_{k}}{m_{i}+m_{j}+m_{k}})\smallskip $%
\end{tabular}%
\   \label{Jacobi_K}
\end{equation}%
where $m_{\ell }$ and $\overrightarrow{r}_{\ell }$ are\ respectively the
mass and the position of the $\ell $-th particle. The coordinate set
associated with the components $H\equiv \Psi _{ij}^{kl}$, is defined by:

\begin{equation}
\begin{tabular}{ll}
$\overrightarrow{x_{ij}}$ & $=\sqrt{2\frac{m_{i}m_{j}}{m_{i}+m_{j}}}(%
\overrightarrow{r}_{j}-\overrightarrow{r}_{i})\smallskip $ \\
$\overrightarrow{y_{kl}}$ & $=\sqrt{2\frac{m_{k}m_{l}}{m_{k}+m_{l}}}(%
\overrightarrow{r}_{l}-\overrightarrow{r}_{k})\smallskip $ \\
$\overrightarrow{z_{ij,kl}}$ & $=\sqrt{2\frac{\left( m_{i}+m_{j}\right)
\left( m_{k}+m_{l}\right) }{m_{i}+m_{j}+m_{k}+m_{l}}}(\frac{m_{k}%
\overrightarrow{r}_{k}+m_{l}\overrightarrow{r}_{l}}{m_{k}+m_{l}}-\frac{m_{i}%
\overrightarrow{r}_{i}+m_{j}\overrightarrow{r}_{j}}{m_{i}+m_{j}})$%
\end{tabular}%
\   \label{Jacobi_H}
\end{equation}

The functions $K$ and $H$ are expanded in the basis of partial
angular momentum, spin and isospin variables, according to:

\begin{equation}
\Phi _{i}(\vec{x}_{i},\vec{y}_{i},\vec{z}_{i})=\sum_{\alpha }\frac{\mathcal{F%
}_{i}^{\alpha }(x_{i},y_{i},z_{i})}{x_{i}y_{i}z_{i}}Y_{i}^{\alpha }(\hat{x}%
_{i},\hat{y}_{i},\hat{z}_{i})  \label{PWB_decomp}
\end{equation}

Here $Y_{i}^{\alpha }(\hat{x}_{i},\hat{y}_{i},\hat{z}_{i})$ generalize
tripolar harmonics containing spin, isospin and angular momentum variables.
Functions $\mathcal{F}_{i}^{\alpha }(x_{i},y_{i},z_{i})$ are so called
partial amplitudes, being continuous in radial variables $x,y$ and $z.$ The
label $\alpha $ represents the set of intermediate quantum numbers defined
in coupling scheme, it includes as well the specification for the type of
FY's component ($K$ or $H$). We have used $j-j$ couplings, represented in
Fig. \ref{Fig_4b_config} [b], and expressed by:

{\small
\begin{equation}
\left[ \left\{ \left( t_{i}t_{j}\right) _{t_{x}}t_{k}\right\} _{T_{3}}t_{l}%
\right] _{\mathcal{T}}\left\langle \left\{ \left[ l_{x}\left(
s_{i}s_{j}\right) _{\sigma _{x}}\right] _{j_{x}}\left[ l_{y}s_{k}\right]
_{j_{y}}\right\} _{J_{3}}\left[ l_{z}s_{l}\right] _{j_{z}}\right\rangle _{%
\mathcal{J}^{\pi }}  \label{K_jj_scheme}
\end{equation}%
} for components of $K$-type, and {\small
\begin{equation}
\left[ \left( t_{i}t_{j}\right) _{t_{x}}\left( t_{k}t_{l}\right) _{t_{y}}%
\right] _{\mathcal{T}}\left\langle \left\{ \left[ l_{x}\left(
s_{i}s_{j}\right) _{\sigma _{x}}\right] _{j_{x}}\left[ l_{y}\left(
s_{k}s_{l}\right) _{\sigma _{y}}\right] _{j_{y}}\right\}
_{j_{xy}}l_{z}\right\rangle _{\mathcal{J}^{\pi }}  \label{H_jj_scheme}
\end{equation}%
} for the $H$-type components. Here $s_{i}$ and $t_{i}$ are the spin and
isospin quantum numbers of the individual particles and $\left( \mathcal{J}%
^{\pi },\mathcal{T}\right) $ are, respectively, the total angular momentum,
parity and isospin of the four-body system. Each amplitude $\mathcal{F}%
_{i}^{\alpha }(x_{i},y_{i},z_{i})$ is thus labelled by the a set
of 12 quantum numbers $\alpha $. The symmetry properties of the
wave function in respect to exchange of two particles impose
additional constraints. One should have $\left( -\right)
^{l_{x}+\sigma _{x}+t_{x}}=\varepsilon $ for
the amplitudes derived from any type of components ($K$ or $H$), while for $%
H-$type amplitudes additional constraint $\left( -\right)
^{l_{y}+\sigma _{y}+t_{y}}=\varepsilon $ is valid as well.  Since
we deal with nucleons (i.e. fermions) Pauli factor $\varepsilon $
is equal -1. The total parity $\pi $ is given by $\left( -\right)
^{l_{x}+l_{y}+l_{z}}$, independently of the coupling scheme in
use.

By projecting each of the Eqs. (\ref{FY1}) on its natural configuration
space basis one obtains a system of coupled integrodifferential equations.
In general one has an infinite number of coupled equations. Note that,
contrary to the 3N\ problem, the number of partial FY amplitudes is infinite even
when the pair interaction is restricted to a finite number of partial waves.
This divergence comes from the existence of additional degree of freedom $%
l_{z}$ in the expansion of the $K$-type components. Therefore we are obliged to make
additional truncations in numerical calculations by taking into account only
the most relevant amplitudes.

\subsection{Boundary conditions}

Equations (\ref{FY1}) are not complete and  should be complemented with the
appropriate boundary conditions. Boundary conditions can be written in the
Dirichlet form. First FY\ amplitudes, for bound as well as for scattering
states, satisfy the regularity conditions:

\begin{equation}
\mathcal{F}_{i}^{\alpha }(0,y_{i},z_{i})=\mathcal{F}_{i}^{\alpha
}(x_{i},0,z_{i})=\mathcal{F}_{i}^{\alpha }(x_{i},y_{i},0)=0  \label{BC_xyz_0}
\end{equation}

For the bound state problem, the wave function is exponentially decreasing
and therefore the
regularity conditions can be completed by forcing the amplitudes $\mathcal{F}%
_{i}^{\alpha }$ to vanish at the borders of the hypercube $\left[ 0,X_{\max }%
\right] \times \left[ 0,Y_{\max }\right] \times \left[ 0,Z_{\max }\right] $,
i.e.:

\begin{equation}
\mathcal{F}_{i}^{\alpha }(X_{\max },y_{i},z_{i})=\mathcal{F}_{i}^{\alpha
}(x_{i},Y_{\max },z_{i})=\mathcal{F}_{i}^{\alpha }(x_{i},y_{i},Z_{\max })=0
\label{BC_BS}
\end{equation}

For the elastic scattering problem the boundary conditions are
implemented by imposing at large values of $z$ the asymptotic
behavior of the solution. In case of N+NNN elastic scattering we
impose at $Z_{\max }$ the solution of the 3N\ problem for all the
quantum numbers, corresponding to the open channel $\alpha _{a}$:

\begin{equation}
\mathcal{F}_{i}^{\alpha _{a}}(x_{i},y_{i},Z_{\max })=f_{i}^{\alpha
_{a}}(x_{i},y_{i})  \label{BC_SS}
\end{equation}

Functions $f_{i}^{\alpha _{a}}(x_{i},y_{i})$ are the Faddeev
amplitudes obtained after solving corresponding 3N\ bound state
problem. Indeed, below the first inelastic threshold, at large
values of $z$, the solution of (\ref{FY1}) factorizes into a bound
state solution of 3N\ Faddeev
equations and a plane wave propagating in $z$ direction with the momentum $%
k_{\alpha _{a}}=\sqrt{\frac{m}{\hbar ^{2}}\left( E_{cm}-E_{3N}\right) }$.
One has:

\begin{equation*}
\mathcal{F}_{i}^{\alpha _{a}}(x_{i},y_{i},z_{i})\sim f_{i}^{\alpha
_{a}}(x_{i},y_{i})\left[ \hat{\jmath}_{l_{z}}(k_{\alpha _{a}}z_{i})+\tan
(\delta )\hat{n}_{l_{z}}(k_{\alpha _{a}}z_{i})\right]
\end{equation*}

There are two different ways to obtain the scattering observables.
The easier one is to extract the scattering phases from the tail
of the solution, namely taking logarithmic derivative of the open
channel's $K$ amplitude $\alpha _{a}$ in the asymptotic region:

\begin{equation}
\tan \delta = \frac{k_{\alpha _{a}}\hat{\jmath}_{l}^{\prime
}(k_{\alpha _{a}}z_{i})-\frac{\partial _{z_{i}}\mathcal{F}_{i}^{\alpha
_{a}}(x_{i},y_{i},z_{i})}{\mathcal{F}_{i}^{\alpha _{a}}(x_{i},y_{i},z_{i})}%
\hat{\jmath}_{l}(k_{\alpha _{a}}z_{i})}{\frac{\partial _{z_{i}}\mathcal{F}%
_{i}^{\alpha _{a}}(x_{i},y_{i},z_{i})}{\mathcal{F}_{i}^{\alpha
_{a}}(x_{i},y_{i},z_{i})}\hat{n}_{l}(k_{\alpha _{a}}z_{i})-k_{\alpha _{a}}%
\hat{n}_{l}^{\prime }(k_{\alpha _{a}}z_{i})}
\end{equation}

This result can be independently verified by using integral representation
of the phase shifts

\begin{equation}
\sin \delta =-\frac{m}{\hslash ^{2}}\int \Phi _{\alpha _{a}}^{(123)}(\vec{x}_{i},\vec{y}_{i})\hat{%
\jmath}_{l}(k_{\alpha _{a}}z)(V_{14}+V_{24}+V_{34})\Psi(\vec{x}_{i},\vec{y}_{i},\vec{z}_{i}) dV.
\label{Eq:Integ_rep_sc_l}
\end{equation}%
Here $\Phi _{\alpha _{a}}^{(123)}(\vec{x}_{i},\vec{y}_{i})$ is a
3N bound state wave function composed by particles (1,2,3). This
wave function is considered to be normalized to unity. Asymptotes
of the wave function $\Psi(\vec{x}_{i},\vec{y}_{i},\vec{z}_{i}) $
is considered to have the same normalization as $\Phi _{\alpha
_{a}}^{(123)}(\vec{x}_{i},\vec{y}_{i})$, i.e. it tends to:
\begin{equation}
\Psi (\vec{x}_{i},\vec{y}_{i},\vec{z}_{i})=\Phi _{\alpha _{a}}^{(123)}\left(
\vec{x}_{i},\vec{y}_{i}\right) \left[ \hat{\jmath}_{l_{z}}(k_{\alpha
_{a}}z_{i})+\tan (\delta )\hat{n}_{l_{z}}(k_{\alpha _{a}}z_{i})\right] .
\end{equation}

A detailed discussion on these technical aspects can be found
in \cite{These_Fred_97,These_Rimas_04}.

\subsection{\protect\bigskip Numerical solution}

In order to solve the set of integro-differential equations -- obtained when projecting
eq.(\ref{FY1}) in conjunction with the appropriate boundary conditions into
partial wave basis --   components $\mathcal{F}_{i}^{\alpha }$ are expanded in terms of piecewise
Hermite spline basis.
\[\mathcal{F}_{i}^{\alpha}(x,y,z)= \sum c^{\alpha}_{ijkl} S_j(x)S_k(y)S_l(z)\]

In this way,
integro-differential equations are converted into an equivalent linear algebra
problem with unknown spline expansion coefficients $c^{\alpha}_{ijkl}$  to determine. In case of
bound state problem eigenvalue-eigenvector problem is obtained:

\begin{equation}
A c=EB c,
\end{equation}
where $A$ and $B$ are square matrices, while $E$ and $c$ are respectively
unknown eigenvalue(s) and its eigenvector(s) to determine. In case of
elastic scattering problem, a system of linear algebra equations is obtained:

\begin{equation}
\left[ A-E_{cm}B\right] c=b
\end{equation}%
where $b$ is an inhomogeneous term imposed by the boundary conditions eq.(\ref{BC_SS}).
Numerical methods used for solving these large scale $N\sim10^7$ eigenvalue problems and linear systems
are given in \cite{These_Rimas_04}.

\section{Results and discussion}

We have used in our calculations  four different Doleschall
potentials derived in references
\cite{D_NPA602_96,D_FBS_98,DB_FBS_99,DB_PRC62_00,DB_NPA684_01,DBPP_PRC67_03,D_PRC_04}.
Hereafter, INOY96 denotes the SB+SDA
version of the potential defined in  \cite{D_NPA602_96,D_FBS_98}.
It consists in short-range non local potentials in
$^1$S$_0$ and $^3$SD$_1$ partial waves continued with a local Yukawa tail
outside R=4 fm. The other partial waves are taken from  AV18.
INOY03  denotes the IS version considered in Ref.
\cite{DBPP_PRC67_03}. It is an updated version of INOY96 which has
a smaller non locality range (R=2 fm) and provides a more accurate description of 2N observables.
INOY04 and INOY04' are the two most recent
versions \cite{D_PRC_04} having the same $^1$S$_0$ and $^3$SD$_1$
potentials as INOY03, completed with the newly defined non local
potentials in P- and D- waves. As in the preceding models, higher
partial waves are also taken from AV18.

All the results have been obtained considering equal masses for
neutrons and protons ($m_n=m_p=m$) with ${\frac{\hbar^2}{m}}=41.47$ MeV fm$^2$.
As mentioned in
section \ref{sec_FY_eq} we  have used isospin formalism,
furthermore assuming the total isospin quantum number $\mathcal{T}$ to be conserved.

\subsection{3N system}

We will start with the presentation of our results concerning 3N systems.
Binding energies for $^{3}$H and $^{3}$He nuclei
are summarized  in Table \ref{B_3N}.
In order to control our accuracy, we have included in this table the results of AV18
\cite{AV18_PRC_95} with and without Urbana IX three-nucleon force \cite{UIX}.
One can see that we are in close agreement with the benchmark calculations of Ref. \cite{Bench_the}.
Our results concerning non-local potentials are slightly
different from those given in \cite{DBPP_PRC67_03,D_PRC_04}.
The small deviations ($\approx$ 15 keV) come from isospin breaking effects which were
fully included in Doleschall calculations \cite{DPC}
while, as  discussed in the above section,
they were  only approximately taken into account in ours.
These isospin effects for AV18+UIX have also been evaluated in
\cite{Bench_the} and were found to be of about 5 keV,
i.e. three times less than in non-local models.
In any case, the small differences related to isospin approximation
can not overcast the main achievement of these non local interactions:
the ability to reproduce experimental 3N binding energies without three-nucleon force.
One can also remark from Table \ref{B_3N} that, while the binding energies obtained
with AV18+UIX are in good agreement with the experimental data,
the value of $\Delta B= B_{3H}-B_{3He}$ is better reproduced by non local models.

\begin{table}[h!]
\caption{$^3$H and $^3$He binding energies (in MeV) calculated
with various versions of non-local Doleschall potentials and with
AV18+UIX  model. Results are compared to experimental values and
previous calculations.
$\Delta B$ denotes the difference between $^3$H and $^3$He binding energies}\label{B_3N}%
\begin{ruledtabular}
\begin{tabular}{l|ll|ll|ll}
 & \multicolumn{2}{c|}{$^{3}$H} & \multicolumn{2}{c|}{$^{3}$He}& \multicolumn{2}{c}{$\triangle$B}\\\hline
        &this work& other                      &this work& other                     &this work& other   \\
INOY96  & 8.556   &                            & 7.882   &                           &0.674    &       \\
INOY03  & 8.497   & 8.482 \cite{DBPP_PRC67_03} & 7.734   & 7.718 \cite{DBPP_PRC67_03}&0.763    & 0.764 \\
INOY04  & 8.476   &                            & 7.711   &                           &0.765    &       \\
INOY04' & 8.464   & 8.481 \cite{D_PRC_04}      & 7.704   & 7.718 \cite{D_PRC_04}     &0.760    & 0.763 \\\hline
AV18    & 7.616   & 7.618(2)\cite{Bench_the}   & 6.914   & 6.917(2)\cite{Bench_the}  &0.699    &\\
AV18+UIX& 8.473   & 8.474(4)\cite{Bench_the}   & 7.739   & 7.742(4)\cite{Bench_the}  &0.734 &\\\hline
Exp. & \multicolumn{2}{l|}{8.482} & \multicolumn{2}{l|}{7.718}  &\multicolumn{2}{l}{0.764}\\ %
\end{tabular}
\end{ruledtabular}
\end{table}

The analysis of the 3N binding energies is shown in Table \ref{TVR_3N}.
One can see that the
major contribution to binding is due to the non-local short range
interaction terms $<V_{nl}>$. Contributions from the local part -- coming
either from long range Yukawa tail or from higher partial waves -- are
marginal in INOY96 model and remain less than 15\% in the other ones.

Comparison between INOYs and AV18+UIX results is instructive. Both models provide
similar energies and rms radii but they result from
values of kinetic and potential energies
which differ as much as $\sim$ 50\%.
The introduction of a non-local interaction reduces the short range repulsion
between nucleons and makes the potential well more shallow.
Deuteron is already obtained with the same binding energy and size as
for AV18 model but its wave function does
not have, at short distances, the sharp slope due to hard-core repulsion.
On the other hand the weaker
$^3$S$_1-^3$D$_1$ coupling in non-local models generates a smaller contribution of
D-state. All these effects reduce the average kinetic energy and favor
stronger binding in the 3N system, which is more compact than deuteron.
If the average size of 3N system between two models is only slightly different,
the average kinetic energy
of nucleons is sensibly smaller in case of non-local interactions.

\begin{table}[h!]
\caption{Expectation values of kinetic ($<T>$) energies, rms radius ($R=\sqrt{<r^2>}$)
and proton radius $r_p$ corresponding to the binding energies of Table \ref{B_3N}.
The values of the potential energy have been separated into contributions coming from the non-local $<V_{nl}>$ and local $<V_{l}>$ terms of the potential.
For AV18+UIX model, $<V_{nl}>$ denotes the contribution of 3NF to potential energy. For $^3$He, the expectation
values of  Coulomb interaction have not been included.}\label{TVR_3N}%
\begin{ruledtabular}
\begin{tabular}{l|lccccc}
&Model   & $<T>$ (MeV) &$-<V_l>$ (MeV)&$-<V_{nl}>$ (MeV) &  $R$ (fm) &  $r_p$ (fm)  \\\hline
$^3$H&INOY96  & 34.24 & 0.776 & 42.02 & 1.656 &  1.561\\
&INOY03  & 33.11 & 5.551 & 36.06 &   1.664  & 1.566\\
&INOY04  & 33.01 & 5.564 & 35.92 &   1.667  & 1.567\\
&INOY04'  & 32.97 & 5.547 & 35.89 &   1.668 & 1.568\\\cline{2-7}
&AV18    & 46.71 & 54.32 & - &   1.770      & 1.654\\
&AV18+UIX& 51.28 & 58.69 & 1.140 &   1.684  & 1.584\\\cline{2-7}
& Exp.   &       &       &       &          & 1.60\\\hline\hline
$^3$He&INOY96  & 33.64 & 0.777 & 41.42 &   1.684  & 1.733 \\
&INOY03  & 32.33 & 5.512 & 35.21 &   1.701  &  1.752\\
&INOY04  & 32.24 & 5.510 & 35.08 &   1.703  & 1.755 \\
&INOY04' & 32.20 & 5.525 & 35.05 &   1.704  & 1.756\\\cline{2-7}
&AV18    & 45.68 & 53.30 &  -     &   1.809  & 1.867  \\
&AV18+UIX& 50.22 & 57.60 & 1.095 &   1.716  & 1.767\\\cline{2-7}
& Exp.   &       &       &       &          & 1.77\\
\end{tabular}
\end{ruledtabular}
\end{table}

\bigskip
The relative contributions (algebraic values) of the most relevant
$V_{NN}$ partial waves to $^3$H  potential energy are given in
Table \ref{RCVNNPW_3N}. The column labelled "others" denotes the
contribution of all partial waves not listed in the table. One can
remark the small role of $P$-waves. It was noticed long time ago
that triton binding energy basically depends on $^1$S$_0$  and
$^3$S$_1$-$^3$D$_1$  $NN$-interactions. Indeed, in absence of
tensor force, the 3N ground state would have L=$0$ and
S=$\frac{1}{2}$ as conserved quantum numbers and in this case
$P$-waves would not contribute at all. The $^3$S$_1$-$^3$D$_1$
tensor coupling introduces an $L=2$ admixture in the wave
function. L=1 state appears only in the second order and
contributes less than 0.1 \%. NN $P$-waves start acting only in
the second order as well, which explains their negligible
contribution to 3N binding energy, as shown in Table
\ref{RCVNNPW_3N}. The reduction of the tensor force is also
sizeable: $^3$D$_1$-waves contribution are considerably smaller
for Doleschall interactions than for AV18.

\begin{table}[h]
\caption{Relative contributions of different $V_{NN}$ partial waves to triton potential energy.}\label{RCVNNPW_3N}
\begin{ruledtabular}
\begin{tabular}{l|llllllllll|l}
       & 3S1   & 1S0   & 3D1   & 3P2    & 3P0    & 1D2     & 3P1     & 1P1      & 3D2     & 3D3    & Others  \\\hline
INOY96 & 57.94 & 29.24 & 12.78 & 0.1658 & 0.1506 & 0.05664 & -0.3810 & -0.04528 & 0.04939 & 0.01505& 0.02518 \\
INOY03 & 58.30 & 30.31 & 11.42 & 0.1350 & 0.1892 & 0.04260 & -0.4214 & -0.05317 & 0.06144 & 0.01634& 0.02307 \\
INOY04 & 58.35 & 30.33 & 11.38 & 0.1409 & 0.1579 & 0.04843 & -0.4304 & -0.05495 & 0.06303 & 0.06460& 0.0113  \\
INOY04'& 58.40 & 30.36 & 11.37 & 0.1478 & 0.7854 & 0.05061 & -0.4368 & -0.05711 & 0.06527 & 0.04333& 0.0089  \\\hline
AV18   & 45.00 & 25.30 & 28.97 & 0.4466 & 0.2272 & 0.1840  & -0.3977 & -0.03244 & 0.07952 & 0.08769& 0.1290  \\
\end{tabular}
\end{ruledtabular}
\end{table}

\begin{table}[tbp]
\caption{Neutron-deuteron (nd)  scattering lengths (in
fm) calculated using Doleschall potentials.}\label{tab:a0_nd}%
\begin{ruledtabular}
\begin{tabular}{l|cc}
         & $^2a_{nd}$ (fm)& $^4a_{nd}$ (fm) \\ \hline
INOY96   & 0.448          & 6.34            \\
INOY03       & 0.523      & 6.34            \\
INOY04   & 0.543          & 6.34            \\
INOY04'  & 0.553          & 6.34           \\\hline
AV18     & 1.26           & 6.34           \\
AV18+UIX & 0.595          & 6.34            \\\hline
Exp.     & 0.65$\pm$0.04  & 6.35$\pm$0.02  \\
\end{tabular}
\end{ruledtabular}
\end{table}

\bigskip
The calculated n-d scattering lengths are presented in Table
\ref{tab:a0_nd}. The quartet value ($^4$a), corresponding to
$J^{\pi}=\frac{3}{2}^+$ state, is independent of the interaction model in use,
furthermore being in full agreement with the experimental one. This robustness
is due to the strong Pauli repulsion,  prohibiting two neutrons to get close to
each other. It follows that only $^3$S$_1-^3$D$_1$ V$_{np}$ waves are important
in describing $J^{\pi}=\frac{3}{2}^+$ state and still only through its, well
controlled, long range part. Therefore, this state  does not contain any
off-shell  physics and can be successfully described by any potential model,
provided it reproduces the $J^{\pi}=1^+$ np scattering observables.

The integral representation  of the phase shifts, eq. (\ref{Eq:Integ_rep_sc_l}),
is used  to study the role of different  V$_{NN}$ partial waves.
In Table \ref{tab:3N_SC_IMP} are given the relative contributions  to this integral.
Their sum, in algebraic values, is normalized to 100.
Results for n-d doublet scattering length ($^2$a)
are presented in the upper half of the table and the quartet ones in the lower part.
It can be seen that for $J^{\pi}=\frac{3}{2}^+$, NN waves other than $^{3}$S$_{1}$
contribute by less than 0.1\%, which confirms the statements above.
The situation is different for $^2$a which results mainly from a cancellation
between $^1$S$_0$ and $^3$S$_1$  and  is more sensitive to
higher NN partial waves, showing a deeper impact into the off-shell physics.
Due to the smallness of $^2$a, all the interaction
effects have to be taken into account very accurately.
In particular, its value is very sensitive to the electromagnetic (e.m.)
interaction terms.
The differences between INOY
predictions and the experiment can therefore be caused by the absence of e.m. terms in these models.
On the contrary e.m. corrections were properly included in AV18+UIX results.
In any case the small discrepancy with data has no consequences in phenomenology,
since $^2$a is by one order of magnitude smaller than $^4$a and its relative contribution
to the scattering cross sections is negligible.

\begin{table}[h!]
\caption{Relative contributions of different NN interaction waves in n-d integral scattering lengths.
Doublet value is in the upper half of the table and quartet in the lower}\label{tab:3N_SC_IMP}
\begin{ruledtabular}
\begin{tabular}{l|llllllllll|l}
       & 3S1    & 1S0   & 3D1   & 3P2    & 3P0    & 1D2     & 3P1     & 1P1      & 3D2     & 3D3    & Other.  \\\hline
INOY96 & -685.3 & 800.3 & -20.14 & 2.664 & -1.240  & -5.097 & -11.97 & 20.07 &-0.7484  &0.6967& 0.7998 \\
INOY03 & -572.6 & 685.2 & -16.73 & 1.955 &  0.2399 & -4.523 & -10.95 & 16.73 &-0.6048  &0.3972& 0.8616 \\
INOY04 & -549.6 & 662.4 & -16.28 & 1.997 & -0.3862 & -4.257 & -10.31 & 15.66 &-0.5025  &0.3749& 0.8932 \\
INOY04'& -538.5 & 650.8 & -15.92 & 2.127 & -1.076  & -4.167 & -9.308 & 15.25 &-0.4694  &0.3563& 0.8825 \\\hline
AV18   & -195.5 & 293.5 & -4.283 & 4.779 &  0.6261 & -0.3768 & -5.944 & 6.190 &-0.3316 &0.9574& 0.3281 \\ \hline\hline
INOY96 & 100.1 & -0.0297 & 0.0183 & -0.2708 & -0.5036 & -0.1510 & 0.8177 & 0.0190  & -0.1644 &  0.0076  & 0.0312 \\
INOY03 & 100.0 & -0.0288 & 0.0199 & -0.2694 & -0.4640 & -0.1529 & 0.8180 & 0.0192  & -0.1641 &  0.00742 & 0.0315 \\
INOY04 & 100.1 & -0.0283 & 0.0197 & -0.2707 & -0.4938 & -0.1515 & 0.8343 & 0.0189  & -0.1653 & -0.0006  & 0.0480 \\
INOY04'& 99.98 & -0.0279 & 0.0195 & -0.2675 & -0.4973 & -0.1503 & 0.9043 & 0.0188  & -0.1651 & -0.0004  & 0.0479 \\\hline
AV18   & 100.1 & -0.0400 & 0.0097 & -0.2672 & -0.4870 & -0.2577 & 0.8081 & 0.0252  & -0.1691 &  0.0096 & 0.0361 \\
\end{tabular}
\end{ruledtabular}
\end{table}

\subsection{4N system}

Our results concerning the $\alpha $-particle binding energy are displayed in Table \ref{B_4N}.
Two series of calculations were
performed, including (upper half of the table) and neglecting (lower half)
Coulomb repulsion between protons.
This latter interaction was provided by Argonne group in
their AV18 code \cite{AV18_PRC_95} and takes into account proton finite size effects.

\begin{table}[htbp]
\caption{Binding energy B (in MeV) and rms radius R (in fm) for $^4$He ground state obtained with Doleschall
and AV18+UIX potentials. The lower part contains Coulomb force. Energies presented in the two last lines of the table
respectively for AV18 and AV18+UIX models have been taken from reference \cite{Nogga_alp,Nogga_thesis}, whereas rms radius from reference
\cite{PPWC_PRC64_01}.}\label{B_4N}
\begin{ruledtabular}
\begin{tabular}{l|llll}
Pot.     & $<T>$ & $-<V>$& $B$     & $R$   \\\hline
INOY96   & 72.80 & 103.8 & 31.00  & 1.353   \\
INOY03   & 69.89 & 99.94 & 30.04  & 1.369   \\
INOY04   & 69.49 & 99.41 & 29.91  & 1.372   \\
INOY04'  & 69.46 & 99.36 & 29.88  & 1.372   \\\hline
AV18     & 98.69 & 123.6 & 24.95  & 1.511   \\\hline\hline
Pot.     & $<T>$ & $-<V>$&$-<E>$   & $R$    \\\hline
INOY96   & 72.45 & 102.7 & 30.19  & 1.358   \\
INOY03   & 69.54 & 98.79 & 29.24  & 1.373   \\
INOY04   & 69.14 & 98.62 & 29.11  & 1.377   \\
INOY04'  & 69.11 & 98.19 & 29.09  & 1.376   \\\hline
AV18     & 97.77 & 122.1 & 24.22  & 1.516   \\
         & 97.80 & 122.0 & 24.23  \; \cite{Nogga_alp,Nogga_thesis} &         \\
AV18+UIX & 113.2 & 141.7 & 28.50  \; \cite{Nogga_alp,Nogga_thesis}  &   1.44  \cite{PPWC_PRC64_01}  \\\hline
Exp      &       &       & 28.30  & 1.47 \\
\end{tabular}
\end{ruledtabular}
\end{table}

Calculations have been done by considering isospin averaged pair interaction, i.e.:
\begin{equation*}
V_{t_1t_2}=P_{nn}(t_{1},t_{2})V_{nn}+P_{pp}(t_{1},t_{2})V_{pp}+P_{np}(t_{1},t_{2})V_{np}
\end{equation*}
where $(t_1,t_2)$ are the isospin quantum numbers of FY amplitudes in equation
(\ref{PWB_decomp}).
They respectively represent $(t_x,T_3)$ for $K$-type amplitudes and $(t_x,t_y)$ for $H$-type.
$P_{nn},P_{pp}$ and $P_{np}$  are the probabilities of finding respectively nn, pp and np pairs in a given isospin state.
Note that, since the number
of protons and neutrons in $\alpha $-particle is equal, one has $P_{nn}(t_1,t_2)=P_{pp}(t_1,t_2)$.

As in 3N\ calculations, we have neglected isospin breaking
effects, considering $\alpha $-particle as a pure $\mathcal{T}=0$
state. Contributions of $\mathcal{T}=1,2$ admixture were
calculated for AV18\ and AV18+UIX models in \cite{Nogga_alp} and
found to be as small as 10 keV. Results for 3N system presented in
last section showed that Doleschall non-local models are more
sensible to isospin breaking: they account for $\approx$15 keV in
$^3$H compared to $\approx$5 keV in AV18+UIX \cite{Bench_the}. In
any case, for the alpha particle these effects should not exceed
some 50 keV and will not affect the physics discussed below.
Notice also that Coulomb corrections obtained by non-local models
exceed  by 70 keV those obtained by AV18, due to the different rms
radii they give.


\bigskip
As mentioned in section (\ref{sec_FY_eq}),
FY calculations have been performed in the $j-j$ coupling scheme.
The following truncations in the partial wave expansion of amplitudes were used:
{\it (i)} V$_{NN}$ waves limited to $l_x\le3$ but always including
tensor-coupled partners, i.e. involving the set
$^1S_0,^3SD_1$, $^1P_1,^3P_0,^3PF_2,^3P_1$, $^1D_2,^3DG_3,^3D_2$,
$^1F_3,^3FH_4,^3F_3$
and {\it (ii)} $l_x+l_y+l_z\le10$.

Convergence was studied as a function of $j_{yz}$=max($j_y$,$j_z$)
for K-like components and $j_{yz}$=max($j_y$,$l_z)$ for H-like,
starting with $j_{yz}\le1$. In Table \ref{tab:alpha_conv} we
present the $\alpha$-particle binding energy results  for INOY04'
and AV18 models respectively. The convergence is rather smooth,
except when passing from $j_{yz}\le5$ to $j_{yz}\le6$. We think
this is an artefact of our truncation procedure which keeps fixed
the basis set in the x-coordinate. Note that the agreement between
our results for AV18 potential and those, given in reference \cite{Nogga_alp,Nogga_thesis}, is
very good. From results displayed on Table
\ref{tab:alpha_conv}, as well as from analogous convergence
patterns seen in 3N calculations, we conclude that Doleschall
potentials converge more rapidly than AV18. This is probably due
to their weaker tensor force, resulting  into wave functions with
stronger spherical symmetry.

\begin{table}[h!]
\caption{Results of $\alpha$-particle binding energy (in MeV) for
INOY04' and AV18 models with
V$_{NN}$ interaction limited to $l_x\le3$ (see text)
and partial wave basis limited to $l_x+l_y+l_z\le10$.
The convergence was searched as a function of $j_{yz}$=max($j_y$,$j_z$)
for K-like components and $j_{yz}$=max($j_y$,$j_z)$ for H-like.
Last line, denoted by *, contains additional calculations
with NN interaction waves up to $j_x\le 6$ and
$l_x+l_y+l_z\le12$.}\label{tab:alpha_conv}
\begin{ruledtabular}
\begin{tabular}{l|ll}
$j_{yz}$& INOY04' &  AV18  \\\hline
 1      &  28.094 &        \\
 2      &  28.661 &        \\
 3      &  28.967 &        \\
 4      &  28.971 &  23.897 \\
 5      &  28.974 &  23.920 \\
 6      &  29.084 &  24.233 \\
 7      &  29.085 &  24.226 \\
 *      &  29.085 &  24.223 \\
\end{tabular}
\end{ruledtabular}
\end{table}


To our opinion the main conclusion of the results displayed in
Table \ref{B_4N} is the possibility offered by the INOY models to
provide a satisfactory description of  A=4 nuclei in terms of
two-body forces alone, as it is already the case for A=2 and 3.
One can argue that this agreement is not yet fully realized in
their present version, for they all slightly overbind the
experimental value: the most favorable version (INOY04') still
exceeds by 0.79 MeV the $^4$He binding energy. One should however
remark that this result is obtained without adjusting any
additional parameter with respect to A=3. On the other hand, the
difference between INOY96 and INOY04' -- due essentially to
different parameterizations of their non local short range parts
-- is  1.1 MeV. It seems thus possible,  by a finer tuning, to
reach an even more precise description of A=4 in a next generation
of potentials. If they are not contradicted by other aspects of
the phenomenology, INOY models offer an alternative description
permitting to avoid three-nucleon forces.

\begin{figure}[htbp]
\begin{center}\mbox{\epsfxsize=14.cm\epsffile{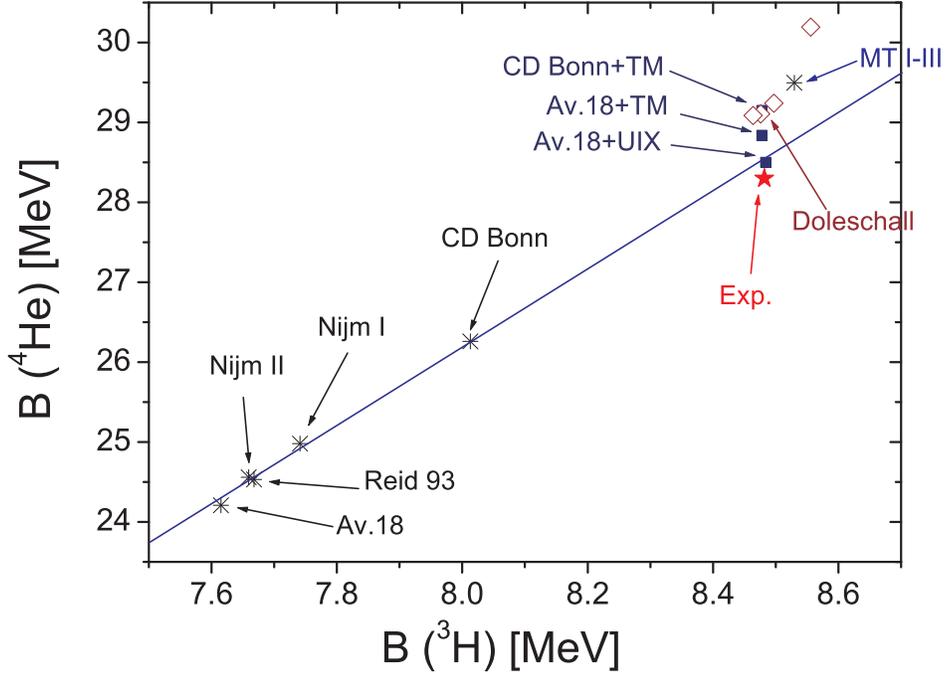}}\end{center}
\vspace{-1cm}
\caption{"(Color online)" Tjon-line for the local and non local NN potentials.}\label{Tjon_line}
\end{figure}

\bigskip
Results of Table \ref{B_4N} have been gathered  in a Tjon plot --
see Figure \ref{Tjon_line} -- which displays the correlation
between $^3$H and $^4$He binding energies for various NN
potentials. One can see  that, due to the small overbinding of
$\alpha$-particle, INOY results (diamond symbols) are outside the
line formed by realistic local  model predictions and, except for
INOY96, are almost superimposed to CD Bonn + TM value.

INOY03, INOY04 and INOY04'
models, which differ in their P-wave structure, give very close results while INOY96,
which has a different non-local S-wave structure, falls out further apart.
This indicates that the S-waves are the key point in binding $\alpha$-particle
and in order to improve the agreement with the experimental value
a better tuning in the  $^3$S$_1$-$^3$D$_1$ and $^1$S$_0$ could be helpful.

\bigskip

Proton rms radii predicted by INOY potentials deserve some comments.
One can see already in 3N systems (see Tab. \ref{TVR_3N})
that they are slightly smaller than the experimental ones. For $\alpha$-particle we
have only calculated average rms of nucleons, without making distinction between neutrons and
protons. The real value of protons rms should be slightly larger.
However, since Coulomb interaction has
very small effect on the alpha particle wave function (rms radii calculated by
taking Coulomb interaction
into and without account  differ only in fourth digit) this value can not differ by more than 0.5\%.
One can therefore see that protons rms provided by INOY model are already by 6\% smaller than the
experimental one, compared to 1-2\% in 3N complex. This fact clearly demonstrates that INOY interactions
are too soft, resulting into a faster condensation of the nuclear matter.
In order to improve the agreement  one should try
to increase the short range repulsion between the nucleons.
This would inevitably  imply a reduction of
3N binding energies, although this reduction is not necessarily very large.
If one chooses the ITF
$^3$S$_1$-$^3$D$_1$ and ISA $^1$S$_0$ potential versions from \cite{DB_NPA684_01}, which give the smallest
probabilities to find nucleons close to each other in 2N systems, the resulting triton underbinding
will be of only $\sim$50 keV.

\begin{figure}[h!]
\vspace{-0.5cm}
\begin{center}\mbox{\epsfxsize=8.cm\epsffile{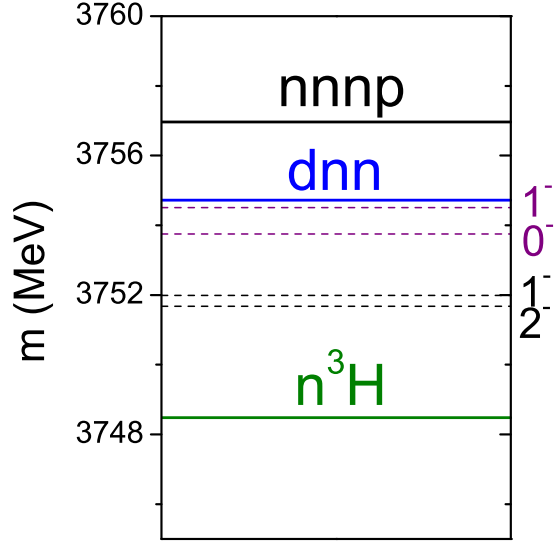}}\end{center}
\vspace{-.5cm}
\caption{"(Color online)" Negative parity resonance above the n+t threshold}\label{nt_en_struct}
\end{figure}

\bigskip
Let us finally consider the  n+$^{3}$H elastic scattering. This
is, in principle, the simplest 4N reaction, being almost a pure
$\mathcal{T}=1$ isospin state and free of Coulomb interaction.
However, its  simplicity is only apparent. In fact, n+$^3$H is a
system with very large  neutron excess  ($\eta={N-Z\over A}
=0.5$). The only stable nucleus having a neutron excess equally
large is $^{8}$He. Let us remind that AV18+UIX  hamiltonian, faces
increasing difficulty when describing neutron rich nuclei. The
more elaborate 3NF model, namely Illinois \cite{PPWC_PRC64_01},
even though being able to improve the agreement with experimental
data, still suffers from similar discrepancies
\cite{These_Rimas_04}. In addition, the n+$^3$H system contains
several nearthreshold resonances of negative parity (see Fig
\ref{nt_en_struct}), which strongly affect the scattering
observables near $E_{cm}\approx3$ MeV. Their internal dynamics is
richer than for bound states and it is not clear that they can be
described by using the same recipes than the one used for solving
the underbinding problem. The  description of the n+$^{3}$H  cross
sections in the resonance region is therefore a very challenging
task for nucleon-nucleon interaction models.

\begin{table}[h]
\caption{n+$^3$H singlet (J$^{\pi}$=0$^+$),
triplet (J$^{\pi}$=1$^+$) and coherent scattering lengths (in fm) along with zero energy cross
sections. Different model results are compared with the experimental data.}\label{tab:n_3H_scatt_len}%
\begin{ruledtabular}
\begin{tabular}{l|llll}
Pot.      & a$_{0^+}$ (fm) & a$_{1^+}$ (fm)& a$_c$ (fm)& $\sigma$ (fm$^2$) \\ \hline
INOY04    & 4.00      & 3.52 & 3.64 & 166.5  \\
INOY04'   & 4.00           & 3.52 & 3.64 & 166.8       \\\hline
AV18      & 4.27           & 3.71 & 3.85 &  187.0 \\
AV18+UIX  & 4.04           & 3.60 & 3.71 &  173.4 \\ \hline
Experimental& 3.70$\pm$0.62 & 3.70$\pm$0.21 & 3.82$\pm$0.07  \cite{HRCK_ZPA_81}  & 170$\pm$3\cite{PBS_PRC_80}\\
          & 4.98$\pm$0.29 & 3.13$\pm$0.11 & 3.59$\pm$0.02   \cite{RTWW_PLB_85}   &      \\
          & 2.10$\pm$0.31 & 4.05$\pm$0.09 &                                      &  \\
          & 4.45$\pm$0.10 & 3.32$\pm$0.02 & 3.607$\pm$0.017 \cite{HDSBP_PRC_90}  &  \\
\end{tabular}
\end{ruledtabular}
\end{table}

We have performed extensive calculations of the n+$^{3}$H scattering states
only by using the INOY04 potential.
The model dependency of the results was checked at $E_{cm}$=0 and 3 MeV with INOY04'.
We present in Table \ref{tab:n_3H_scatt_len} the calculated  singlet ($a_{0^+}$) and triplet
($a_{1^+}$) scattering lengths together with the deduced coherent value
\begin{equation}
a_{c}=\frac{1}{4} \left( a_{0^+} + 3a_{1^+} \right)  \label{eq_coherent_sl}
\end{equation}%
and the zero energy cross section
\begin{equation}
\sigma (0)=\pi \left( a_{0^{+}}^{2}+3a_{1^{+}}^{2}\right)
\label{eq_zero_en_crss}
\end{equation}
Results for AV18 and AV18+UIX models have also been obtained
and agree at 1\% level with those given in Ref. \cite{VRK_98}.

The $J^{\pi }=0^{+}$ and $1^{+}$ positive parity states,
determining the low energy behavior of the n+$^{3}$H  cross
section, do not have any $S$-matrix singularity, except the triton
bound state threshold. It is therefore not surprising that the
n+$^{3}$H scattering lengths are found to be correlated with 3N
binding energy, in a similar way as $n+d$ doublet scattering
length is \cite{Philips}. This is the reason why realistic local
interaction models, providing too low 3N binding energies,
overestimate n+$^{3}$H zero energy cross sections. Once triton
binding energy is corrected, for instance by implementing 3NF, a
value close to the experimental one is automatically obtained.
From Table \ref{tab:n_3H_scatt_len} it can be seen that Doleschall
potential agrees with the lower bound of experimentally measured
zero energy cross section, whereas AV18+UIX model coincides with
its upper bound. The zero energy scattering cross section is thus
fairly well reproduced.

\begin{figure}[h!]
\vspace{-0.5cm}
\begin{center}\mbox{\epsfxsize=16.cm\epsffile{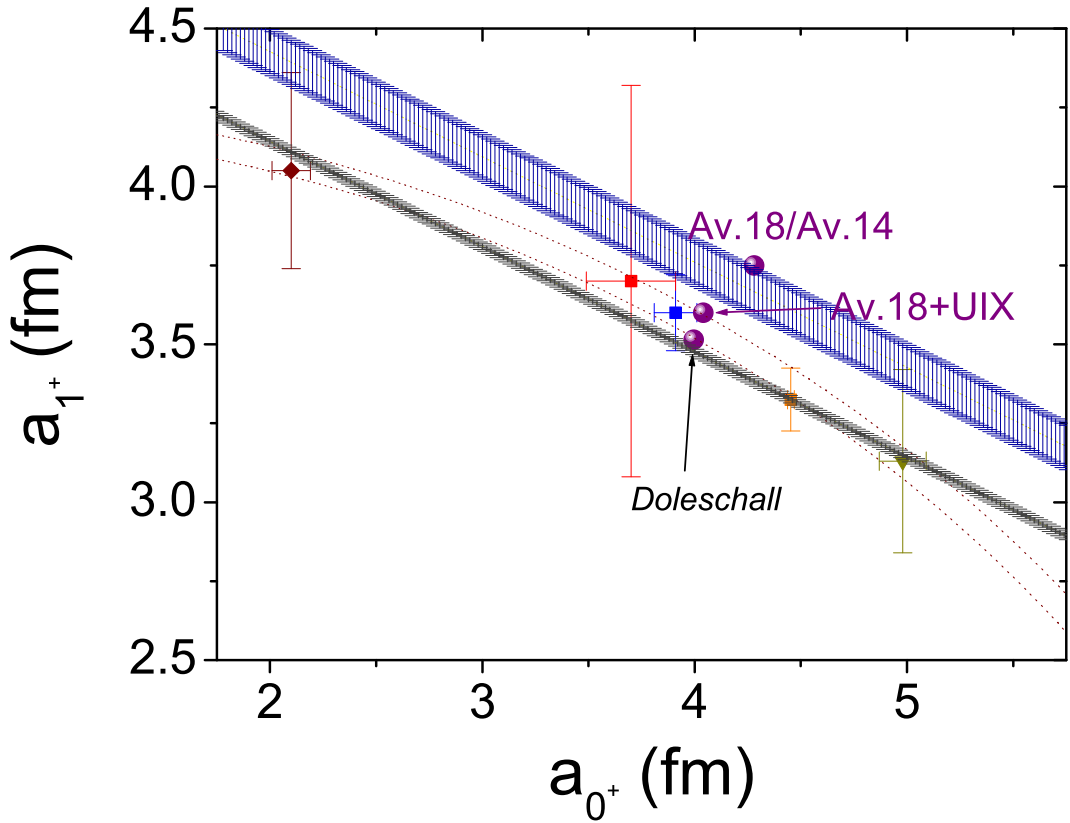}}\end{center}
\vspace{-1cm}
\caption{"(Color online)" Extraction procedure for n+$^{3} $H singlet ($a_{0^{+}}$) and
triplet ($a_{1^{+}}$) scattering lengths
from measurements of zero energy cross section (elliptic band) \cite{PBS_PRC_80}
and coherent scattering length (linear bands) \cite{HRCK_ZPA_81,RTWW_PLB_85,HDSBP_PRC_90}.
The values of $a_i$ are given by the intersection of these two curves.
Band widths are related to experimental errors and, even being small, they make their
determination very unstable.
}\label{Fig_uncert_slen_nt}
\end{figure}

The situation with scattering lengths looks more precarious.
The values found in literature  are hardly compatible with each other \cite{C_NPA684_01}
as it can be seen in Table \ref{tab:n_3H_scatt_len}.
The usual way to get $a_{i}$ is to express them in terms of the measured quantities $a_c$ and
$\sigma(0)$, by reversing relations  (\ref{eq_coherent_sl}) and  (\ref{eq_zero_en_crss}).
This procedure, represented in Fig. \ref{Fig_uncert_slen_nt}, is numerically unstable.
Indeed, once $\sigma (0)$ is fixed, the domain of permitted  $a_{1^+}$ and $%
a_{0^+}$ values is given  by the ellipse of eq. (\ref{eq_zero_en_crss}) in the $(a_{0^+},a_{1^+})$ plane.
Since there are uncertainties in $\sigma (0)$, the permitted
values of scattering lengths are  trapped in-between two ellipsis
(dot curves in Fig. \ref{Fig_uncert_slen_nt}).
On the other hand, each measurement of $a_{c}$ restricts $a_{1^{+}}$
and $a_{0^{+}}$  values to lie on a straight line
which spreads into a band due to experimental errors (see Figure). The
lower band displayed in Figure  \ref{Fig_uncert_slen_nt}
follows from the R-matrix analysis result  $a_{c}=3.607\pm 0.017$ $fm$  \cite{HDSBP_PRC_90},
while the upper one comes from the experimental measurement $a_{c}=3.82\pm 0.07$ fm from \cite{HRCK_ZPA_81}.
By assuming an exact value of $a_{c}$, e.g. $a_c=3.624$ $fm$ given by the top of the lower band,
the present -- though small -- experimental error in $\sigma (0)$ leads to  two
sets of solutions which spread over a wide range:
{\it (i)}  $a_{0^{+}}=[4.31-5.00]$, $a_{1^{+}}=[3.16-3.40]$ and
{\it (ii)} $a_{0^{+}}=[2.25-2.94]$, $a_{1^{+}}=[3.85-4.08]$ fm.
This example illustrates the difficulty of extracting reliable
values of $a_{0^{+}}$ and $a_{1^{+}}$.
The accurate determination of $a_i$ would
require to gain one order of magnitude in measuring both $\sigma (0)$ and
$a_c$.

As it can be seen also from Figure \ref{Fig_uncert_slen_nt},
the coherent scattering length  value $a_{c}=3.82\pm 0.07$ fm of
reference \cite{HRCK_ZPA_81} is in evident disagreement with the experimentally measured zero
energy cross sections, since it does not intersect $\sigma (0)$ ellipsis.
In this respect, the more recent values $a_{c}=3.607\pm 0.017$ fm \cite{HDSBP_PRC_90}
and $a_{c}=3.59\pm 0.02$ fm \cite{RTWW_PLB_85} are more reliable.
Doleschall non-local
potential provides $a_{c}=3.63$ fm, one standard deviation from these
measurements, and seems to be more compatible with data than AV18+UIX model.
Fig. \ref{Fig_uncert_slen_nt} suggests also
that the real value of the zero energy cross section should coincide with the lower bound
of the experimental result.

\begin{figure}[h!]
\begin{center}\mbox{\epsfxsize=14.cm\epsffile{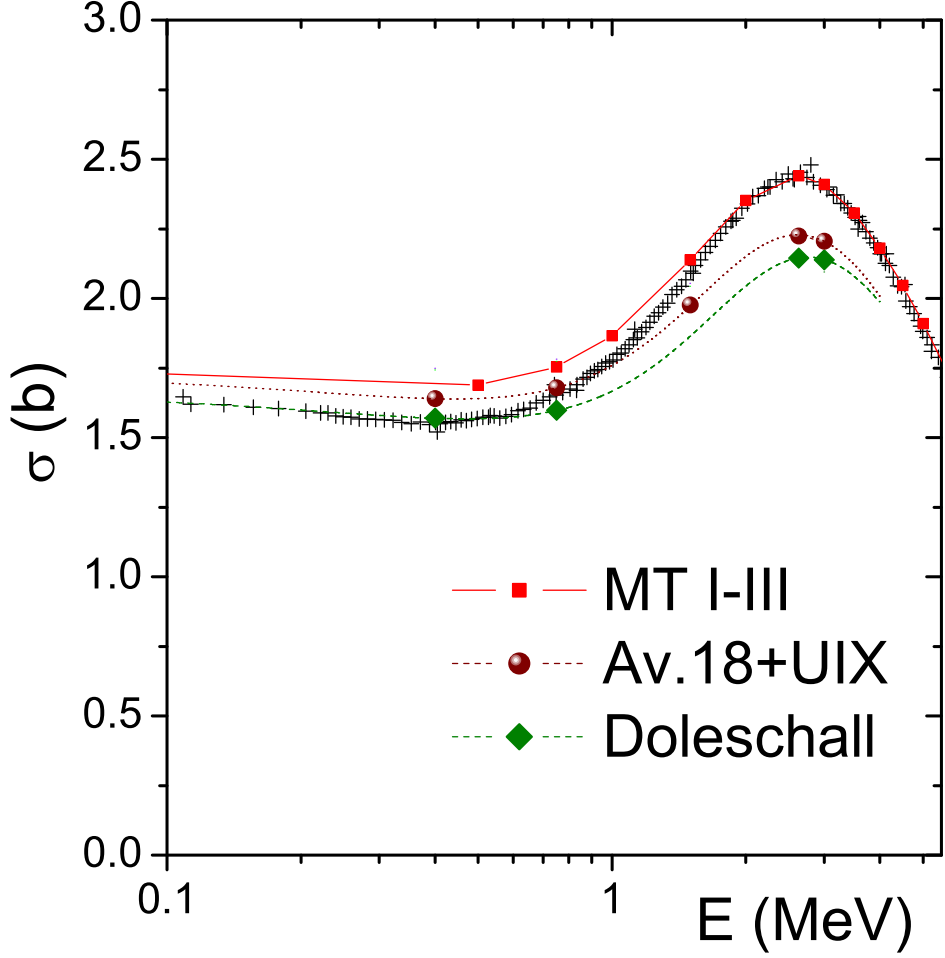}}\end{center}
\vspace{-1.0cm}
\caption{"(Color online)" Comparison between experimental and
theoretical n-$^3$H total cross section calculated with several
local and non-local NN potentials.}\label{sig_nt}
\end{figure}

\bigskip
The success in describing n+$^{3}$H scattering lengths by
Doleschall potential is visible at slightly higher energies as
well. In Figure \ref{sig_nt} we present our calculated  elastic cross section
for the scattering energies in the n+$^{3}$H center of
mass energy  range from 0 to 3 MeV. Doleschall potential reproduces
experimental cross sections near its minima at E$_{cm}\approx 0.4$
MeV. In this region both MT I-III -- the only potential known to us
being capable to reproduce the resonant region \cite{CC_PRC58_98} -- and
AV18+UIX overestimate the experimental value.

In  previous works
\cite{These_Fred_97,CCG_NPA631_98,CCG_PLB447_99,C_FBS12_00,C_NPA684_01}
we pointed out that local realistic interaction models
underestimate the cross sections near the resonance peak,
$E_{cm}=$3 MeV. At that time, calculations had been however
performed with a limited number of partial waves and the failure
was attributed in Ref. \cite{F_PRL83_99} to a lack of convergence.
Recently we have  considerably increased our basis set and have
shown that the disagreement is indeed a consequence of nuclear
models \cite{These_Rimas_04,LC_TUNL}.  The convergence of our
present results is shown in Table \ref{tab:nt_conv}, following the
same truncation criteria as for $\alpha$-particle (see Table
\ref{tab:alpha_conv}). The number of FY partial amplitudes
involved in n+t scattering calculations is considerably larger
than for a pure $0^+$ bound state and  we have not been able to go
in the PWB as far than in Table \ref{tab:alpha_conv}. One can
however remark that results displayed on Table \ref{tab:nt_conv}
converge pretty well, and provide at least three digit accuracy.

\begin{table}[h!]
\caption{Convergence of n+t scattering lengths and selected
phaseshifts at E$_{cm}$=2.625 MeV for INOY04 model. Corresponding
mixing parameters are given in parentheses.}\label{tab:nt_conv}
\begin{ruledtabular}
\begin{tabular}{l|llcc}
$j_{yz}$& $a_0^+$  (fm)& $a_1^+$ (fm) &  $\delta(1^+)$   (deg)          &  $\delta(1^-)$  (deg)   \\\hline
 1      &   3.889 &  3.609  &    ---                  &  ---                  \\
 2      &   3.995 &  3.508  & -56.13  -0.757 (0.803)  & 21.32   39.54 (-42.05)\\
 3      &   3.995 &  3.513  & -56.12  -0.759 (0.803)  & 21.39   39.74 (-43.06)\\
 4      &   3.995 &  3.515  & -56.12  -0.759 (0.803)  & 21.39   39.76 (-43.16)\\
\end{tabular}
\end{ruledtabular}
\end{table}

Implementation of 3NF is just able to improve zero-energy
cross sections and is not efficient at the resonance energies.
Doleschall non-local potentials seem to suffer from a similar defect:
the phase shifts obtained using INOY04  are even slightly
smaller in their absolute value (except for 2$^{-}$ state) than those obtained
with local potentials and the total cross section is slightly worse.
In fact the reduction of positive-parity phase shifts is a consequence
of improving triton binding energies. As was
previously discussed, n+$^{3}$H scattering lengths are linearly correlated with the
triton binding energy.
Whatever the way one uses to increase
triton binding, by means of non-local interaction or 3NF, the final result
will inevitably be a reduction of 0$^{+}$ and 1$^{+}$ n+$^{3}$H scattering
lengths and low energy phase shifts (in absolute value).
It turns out that by the same way, we reduce in absolute value the 0$^{-}$ and 1$^{-}$  phase
shifts.
Only 2$^{-}$ phases are slightly increased in both AV18+UIX and Doleschall
models. Thus we have a real puzzle for the interaction models: on one hand
they have to reduce low energy cross sections, while on the other hand cross
sections in the resonance region should be significantly increased.
The fact that all realistic interactions systematically
suffer in the resonance region let us believe that the underlying reason of this
disagreement is not related to the non-locality or to 3NF effects.
The observed discrepancies have different background than the underbinding problem.
Where does this failure comes from?

When analyzing n+$^{3}$H cross sections in the resonance region, one
should first recall their origin. They are negative parity states
and their symmetry is consequently different from the positive parity ones,
which are dominated by S-waves.
The good agreement of the scattering lengths provided by Doleschall potential
as well as its success in reproducing the low energy cross
section minima, makes us believe that the positive parity phases are quite
well reproduced in the resonance region as well.
On the other hand, negative parity phase shifts
should be rather far from reality, causing a disagreement with the experimental data.

\begin{table}[h]
\caption{Relative contributions of different NN interaction waves in n-$^{3}$H
integral scattering lengths (second half of the table).}\label{tab:4N_V_IMP}%
\begin{ruledtabular}
\begin{tabular}{c|cc|cccccccccc|c}
       &$J^\pi$&$E_{cm}$ (MeV) & 1S0   & 3S1   &   1P1   & 3P0    &  3P1   & 3P2     & 1D2     & 3D1    & 3D2    & 3D3       & others  \\\hline
INOY04'& 0+  &0.0     & 75.95 & 22.83 &  3.588  & -1.301 & 1.942  & -0.8112 & -0.8661 & -1.433 & 0.1145 &  0.006131 & 0.0188 \\
INOY04'& 0+  &3.0     & 79.79 & 19.68 &  3.413  & -1.046 & 1.913  & -0.5796 & -0.5796 & -1.837 & 0.0745 &  0.005113 & -0.0803 \\\hline
INOY04 & 0-  &3.0     & 61.93 & 67.82 & -0.3569 & 32.97  & -26.58 & 2.190   &  1.897  & -40.41 & 0.6489 & -0.6528   & 0.5551 \\
INOY04'& 0-  &3.0     & 64.16 & 70.14 & -0.3769 & 32.83  & -29.71 & 2.305   &  1.987  & -41.90 & 0.6723 & -0.6919   & 0.5750 \\
INOY04'& 2-  &3.0     & 39.75 & 54.90 & -0.1640 & 1.063  & -6.893 & 14.29   &  0.0970 & -3.182 & 0.1476 &  .0001    & 0.0092 \\
\end{tabular}
\end{ruledtabular}
\end{table}

To understand the possible source of such a disagreement, we have
calculated -- as for nd case -- the relative contribution of the different NN partial waves in the
integral expression of the phase shifts (\ref{Eq:Integ_rep_sc_l}).
The obtained results are summarized in Table \ref{tab:4N_V_IMP}.
First two rows correspond to the 0$^+$ at E=0 and E=3 MeV.
One can see that positive parity
states are completely controlled by the interaction in $^3$SD$_1$ and $^1$S$_0$
waves, at zero energies as well as at E$_{cm}$=3 MeV, close to
resonance peak. The role of higher partial waves is marginal.
The situation changes dramatically in negative parity states (values on 3-5 rows).
Contribution of P-wave interactions becomes comparable to S-wave.
On the other hand, the  non triviality of physics in the resonance region is reflected
by the strong compensation of different P-waves
as well as $^3D_1$ and $^3S_1$
components in $^3$SD$_1$ channel. In addition, different P-waves dominate in different
states: in 2$^-$ state, the $^3$P$_2$ waves are the most relevant, whereas $^3$P$_0$ is almost
negligible; in 0$^-$ state, the $^3P_0$ wave has the largest contribution, while $3P_2$
fades away.

Finally, we would like to comment that all the observables where
NN P-waves are contributing, have tendency to disagree with the
experimental data. A small disagreement can already be seen in n-d
doublet scattering lengths (Table \ref{tab:a0_nd}), whereas
triton, described by the same quantum numbers  but where P-waves
are negligible, is perfectly reproduced. Other examples could be
the 3N analyzing powers \cite{P_waves_Gloe,P_waves_Pisa}, as well
as increasing discrepancy when describing binding energies of
neutron rich nuclei (see Figure \ref{Fig_He_Exp_V18}). One should
also recall that  P-waves in most of the NN interactions are tuned
on n-p and p-p data. Moreover, p-p P-waves are overcasted by
Coulomb repulsion, while n-n P-waves are not directly controlled
by experiment at all. This study suggests that  CSB and CIB
effects can be sizeable in n-n P-waves and provide a possible
explanation for the disagreement observed in n-$^3$H resonance
region.

\section{Conclusions}

During the last decade, a series of non-local NN potentials has been developed by
Doleschall and collaborators and were found to provide an overall satisfactory
description of the 2N and 3N system.
If they left unsolved some of the theoretical nuclear problems
-- like the so called $A_y$ puzzle  --
they constitute an undoubted success  for their ability
to reproduce the experimental binding energy of $^3$H and $^3$He nuclei without adding three-nucleon forces.

\bigskip
In this work, we have examined the possibilities for these non-local potentials
to describe the 4N system as well. This system is a cornerstone in the
nuclear ab-initio calculations and a crucial test for the nuclear models.
Not only because therein the underbinding  problem manifests in its full
strength, but also because of the rich variety of scattering states it possesses.

\bigskip
We have found that non-local NN models,
could well provide 3N and 4N binding energies in agreement with the experimental data
without making explicit use of three-nucleon forces.
They offer an alternative solution to cope with the nuclear underbinding problem,
than the one offered by the local plus 3NF philosophy.

In their present form, they overbind $^4$He
by some 0.7 MeV, a discrepancy much smaller that all the existing
models and which reverses the lack of binding observed in  most realistic potentials.
The fluctuations in the predictions between different versions
of the non-local models suggest that a finer parametrization of
their internal non-local part could be enough to make them fully successful in that point.
The current version seems to be too soft in the short range region,
thus giving slightly too small rms radii of  light nuclei.

\bigskip
By calculating  n+$^{3}$H scattering states, we have found that Doleschall
models are also very encouraging in describing the low energy parameters,
providing an even better description than Nijm II or
AV18+UIX. However they fail also in reproducing the elastic cross section few
MeV above, in the resonance region. In a series of preceding works
\cite{CCG_NPA631_98,CCG_PLB447_99,LC_TUNL,These_Rimas_04} we have shown that local realistic
interactions, even implemented with 3NF, underestimate the cross sections at the
resonance peak E$_{\rm cm}$= 3 MeV.
Unfortunately, the non local models do not solve this problem.
On the contrary, they even provide slightly smaller values of the cross section.
We believe than the reason of this failure is common to all realistic
models and lies in the nucleon-nucleon P-waves themselves.
The analysis of their contribution shows that, contrary to $^4$He  binding energy, they
play a crucial role in the n+$^{3}$H  cross sections. If the n-p P-waves seem to be
well controlled by the experimental data, one still has a relative freedom in
the n-n ones to improve the description.

\bigskip
{\bf Acknowledgements:}
Numerical calculations were performed at
Institut du D\'eveloppement et des Ressources en Informatique Scientifique (IDRIS) from  CNRS
and at Centre de Calcul Recherche et Technologie (CCRT) from CEA Bruy\`eres le Ch\^atel.
We are grateful to the staff members of these two organizations for their kind hospitality and useful advices.

\bibliographystyle{plain}
\bibliography{apssamp}

\end{document}